# Case Study for Running Memory-Bound Kernels on RISC-V CPUs

Valentin Volokitin[1], Evgeny Kozinov[1], Valentina Kustikova[1], Alexey Liniov[1], Iosif Meyerov[1*]

[1] Lobachevsky State University of Nizhni Novgorod, 603950 Nizhni Novgorod, Russia
meerov@vmk.unn.ru

**Abstract.** The emergence of a new, open, and free instruction set architecture, RISC-V, has heralded a new era in microprocessor architectures. Starting with low-power, low-performance prototypes, the RISC-V community has a good chance of moving towards fully functional high-end microprocessors suitable for high-performance computing. Achieving progress in this direction requires comprehensive development of the software environment, namely operating systems, compilers, mathematical libraries, and approaches to performance analysis and optimization. In this paper, we analyze the performance of two available RISC-V devices when executing three memory-bound applications: a widely used STREAM benchmark, an in-place dense matrix transposition algorithm, and a Gaussian Blur algorithm. We show that, compared to x86 and ARM CPUs, RISC-V devices are still expected to be inferior in terms of computation time but are very good in resource utilization. We also demonstrate that well-developed memory optimization techniques for x86 CPUs improve the performance on RISC-V CPUs. Overall, the paper shows the potential of RISC-V as an alternative architecture for high-performance computing.

**Keywords:** High-Performance Computing · RISC-V · ISA · C++ · Performance Analysis and Optimization · Memory-Bound Applications

## 1 Introduction

The development of new CPU architectures significantly affects the progress of computer technologies and their application in computer-aided design and engineering. At first glance, there has already been a variety of CPU architectures that satisfy many current needs. These architectures have gone through a thorny path from the first ideas and experimental samples to full-fledged products mass-produced by leading microprocessor manufacturers. It is also necessary to take into account the difficulties with the deployment of new devices, the development of a full stack of specific software and growing of educational ecosystem. All these things are quite expensive and very difficult. However, attempts to freeze progress and settle for only incremental improvements will by no means lead to any significant breakthroughs that people need. Additionally, the closeness and commercial ownership of the existing proprietary architectures (x86, ARM, and others) leads to complicated problems and limita-tions. Technologies controlled by large companies are usually closed to change, which reduces the potential for further development.

The project of a new free and open architecture RISC-V [1, 2] based on the RISC (Reduced Instruction Set Computer) concept [3] which was presented more than 10 years ago at the University of California at Berkeley deserves attention. In just 12 years hardware and software developers have managed to introduce quite efficient CPUs, publicly available for purchase and use. The performance of existing RISC-V CPUs is still far even from mobile x86 and ARM CPUs, but progress in this area is proceeding at a significant pace. It is unlikely that anyone dares to predict when the first high-performance RISC-V CPU will be created, but the prospects look quite real, and it is confirmed by the current announcements of developers, the investments of industry leaders (for example, Intel), and the growing interest of the community [4].

In this paper, we analyze the performance of two available RISC-V devices in solving problems in which memory management is the main factor affecting computation time. Our main interest is to assess current opportunities and future prospects and answer the following key questions:

- What are the opportunities to adapt existing system software to work on RISC-V CPUs, and what efforts are needed?
- What performance indicators related to the memory subsystem of available RISC-V devices are achievable on standard benchmarks that are commonly used on x86 and ARM architectures?
- How does the attainable performance of RISC-V devices compare with the peak performance specified by the hardware manufacturers?
- Are well-established memory optimization techniques applicable to improve performance on RISC-V devices?

To get the first answers to these questions, we tested performance on two RISC-V devices (Mango Pi MQ-Pro and StarFive VisionFive), an ARM device (Raspberry Pi 4 model B), and a high-performance Intel Xeon 4310T server. We study performance on the following memory-bound benchmarks: a standard STREAM test [5], an in-place dense matrix transpose algorithm, and an image filtering algorithm. On the STREAM benchmark, we determined the memory bandwidth for each of the devices. Using the implementations of transposition and filtering algorithms, we studied how well-known techniques for optimizing performance by improving the reuse of data loaded into a cache affect the computation time on RISC-V devices. When analyzing the results, we paid not so much attention to comparing the total computation time, which obviously looks unfair, but to studying how efficient the utilization of available computing resources is. To the best of our knowledge, this work is at least one of the first papers analyzing the performance of algorithms on RISC-V devices.

## 2 RISC-V Architecture

RISC-V is an open instruction-set architecture (ISA) that was originally designed for research and education [1, 2, 6–8]. It is developed from scratch, taking into account the shortcomings of other open architectures and free from the issues of proprietary architectures that are forced to maintain backward compatibility. RISC-V avoids "over-architecting" for a particular microarchitecture style and has a modular design, a compact base ISA and many extensions [2, 8]. So it can be used in systems of any complexity up to high-performance devices like manycore CPUs or accelerators. Support for IEEE-754 floating-point standard, extension for vector operations, privileged and hypervisor architecture allow us to develop both conventional and HPC applications, operating systems, and virtualization software.

Architecture authors develop the corresponding CPU core microarchitectures, processors, and complete systems. Since 2011, when the "Raven 1" SoC was created (ST 28nm FDSOI), they have released a number chips of Raven [9, 10], Hurricane [11, 12], Craft, Eagle, BROOM [13] families. The latter of these use the BOOM microarchitecture [14–16], which is structurally similar and performance competitive with commercial high-performance out-of-order cores. CPU and system-on-a-chip implementations of the architecture are performed by tens of companies and ranged from microcontrollers to high-performance cluster prototype [17]. The number of RISC-V processor cores shipped to date exceeds 10 billions [18].

The RISC-V software stack includes all the necessary tools for application development. The operating system (Linux), compiler (gcc, Clang), core libraries and programming tools are available for every existing implementation. Current prototypes support a limited set of HPC technologies, namely OpenMP, a set of base libraries (openmpi, openBLAS, fftw), and a set of applications which can be compiled and built on RISC-V (WRF, BLAST, GROMACS, VASP, and others) [19, 20]. However, the interest and support from the academic community, commercial companies and government organizations [20–23] will likely bring RISC-V systems to the level of high-performance solutions in the near future.

## 3 Benchmarking Methodology

### 3.1 Infrastructure

We employed two currently available RISC-V devices:

1. Mango Pi MQ-Pro (D1) with Allwinner D1 processor (1 x XuanTie C906, 1GHz) and 1GB DDR3L RAM. Ubuntu 22.10 operating system (RISC-V edition) and GCC 12.2 compiler were installed.

Some architectural features of C906 are as follows: RV64IMAFDCV ISA, 5-stage single-issue in-order execution pipeline, L1 2-way set-associative I-Cache and 4-way set-associative D-Cache with a size of 32 KB each and cache line size of 64 bytes, Sv39 MMU, fully associative L1 uTLB with 20 entries (10 I-uTLB and 10 D-uTLB), 2-way set-associative L2 jTLB of 128 entries, gshare branch predictor with 16 KB BHT, hardware prefetch for instructions (the next consecutive cache line is prefetched) and data (two prefetch methods: forward and backward consecutive and stride-based prefetch with stride less or equal 16 cache lines), 16, 32, 64-bits integer and fp scalar and 512-bit vector operation including fp FMA.

2. StarFive VisionFive (v1) with StarFive JH7100 processor (2 x StarFive U74, 1 GHz) and 8 GB LPDDR4 RAM. OS Ubuntu 22.10 (RISC-V edition) and GCC 12.2 compiler were installed.

Some architectural features of U74 core are as follows: RV64IMAFDCB ISA, 8-stage dual-issue in-order execution pipeline, L1 2-way set-associative I-Cache and 4-way set-associative D-Cache with a size of 32 KB each, cache line size of 64 bytes and a random re-placement policy (RRP), 128 KB 8-way L2 cache with a RRP, hardware data prefetch (forward and backward stride-based prefetch with large strides and automatically increased prefetch distance), Bare and Sv39 MMU modes, fully associative L1 ITLB and DTLB with 40 entries each, direct mapped L2 TLB with 512 entries, branch predictive hardware with 16-entry BTB and 3.6 KB BHT, 64-bits integer and 32, 64-bits floating point scalar operation including fp FMA.

To compare results, we also used a Raspberry Pi device (ARM) and an Intel Xeon server processor (x86) with the following configuration:

1. Raspberry Pi 4 model B with Broadcom BCM2711 (4 x Cortex-A72, up to 1.5 GHz) processor and 4GB LPDDR4 RAM. Ubuntu 20.04 operating system and GCC 9.4 compiler were installed.
2. Server with 2 x Intel Xeon 4310T (2 x 10 Ice Lake cores, up to 3.4 GHz) and 64 GB DDR4 RAM. CentOS 7 operating system and GCC 9.5 compiler were installed. Only the cores of the first CPU were used to eliminate the occurrence of NUMA effects that are obviously absent on other devices.

A direct comparison of powerful server hardware and low-power devices may seem patently disadvantageous for the latter, however, we decided to include the x86 server in the comparison because performance results on a x86 server look reasonable and expected by the HPC community.

### 3.2 Benchmarks

In this paper, we study the performance of RISC-V devices on three memory-bound benchmarks. First, we experimentally determine the memory bandwidth using the commonly applied STREAM benchmark [1], which performs elementary operations on vectors. Measuring memory bandwidth allows us to interpret the results of subsequent experiments on x86, ARM and RISC-V devices and compare them, taking into account the capabilities of the hardware. Next, we consider the in-place matrix transpose algorithm, which is one of the basic dense linear algebra algorithms. We present several implementations of the algorithm and check how typical memory optimization techniques performing well on x86 and ARM CPUs work on RISC-V devices. Finally, we follow the same idea in studying the Gaussian Blur algorithm, successively applying different approaches to code optimization and testing to what extent they speed up calculations on different devices, with a particular focus on RISC-V devices. Summarizing the obtained results, we formulate the main conclusions about the future prospects for using RISC-V devices in HPC applications.

### 3.3 Performance Metrics

The computation time is typically used as the main performance metric. However, we should keep in mind that our comparison involves a single-core low-power processor of the RISC-V architecture, which is still at the beginning of its development, and a 10-core powerful Xeon processor that uses many advances in the field of high-performance computing. Therefore, in addition to the computation time, we also used relative metrics that allow us to make a fair comparison in terms of the utilization of available computational resources.

Given that the RISC-V architecture is still experimental, it is not clear which of the optimization techniques typical for x86 and ARM work well on RISC-V. Therefore, the following question is of interest: what kind of improvement from a naïve version of the algorithm can be obtained by performing a series of memory optimizations typical for conventional CPUs. This metric allows us to understand what kind of improvement can be leaded to by a certain optimization in each particular case, supplementing the computation time, which is dependent from a device features.

Another metric we employ allows us to evaluate how efficiently we use the available memory channels. To evaluate this, we introduce the following metric. At first we calculate the ratio of the data number of bytes that needs to be moved between DRAM and CPU to the computation time. We divide the value calculated in this way by the achieved memory bandwidth, measured by the STREAM benchmark. The result belongs to the segment from zero to one and is dimensionless. The closeness of this value to one indicates that the algorithm uses the bandwidth of the memory channels quite rationally. This metric allows us to compare the devices, taking into account their significantly different performance of the memory subsystem.

Overall, absolute (computation time, memory bandwidth) and relative (speedup from a naive implementation, “utilization of a memory subsystem”) metrics allow us to draw conclusions about the current state of the considered RISC-V devices.

## 4 Numerical Results

### 4.1 STREAM Benchmark

The STREAM benchmark [1] is one of the popular ways to measure achieved memory bandwidth. Like other similar benchmarks, STREAM is based on the idea of reading and writing an array of data from the memory of the corresponding level. STREAM uses 4 tests that have different bytes/iter and FLOPS/iter values:

1. COPY – is simple copying from one array to another (a[i]=b[i]). This operation transfers 16 bytes per iteration and does not perform floating point calculations.
2. SCALE – is copying from one array to another with multiplication by a constant (a[i]=d*b[i]). This operation transfers 16 bytes per iteration and does 1 FLOPS/iter.
3. SUM – the sum of elements from two arrays is stored in the third array (a[i]=b[i]+c[i]). This operation transfers 24 bytes per iteration, but still does 1 FLOPS/iter.

4. TRIAD – FMA (fused multiply-add) from elements from two arrays (a[i]=b[i]+d*c[i]) is stored in the third array. This operation transfers 24 bytes per iteration and does 2 FLOPS/iter.

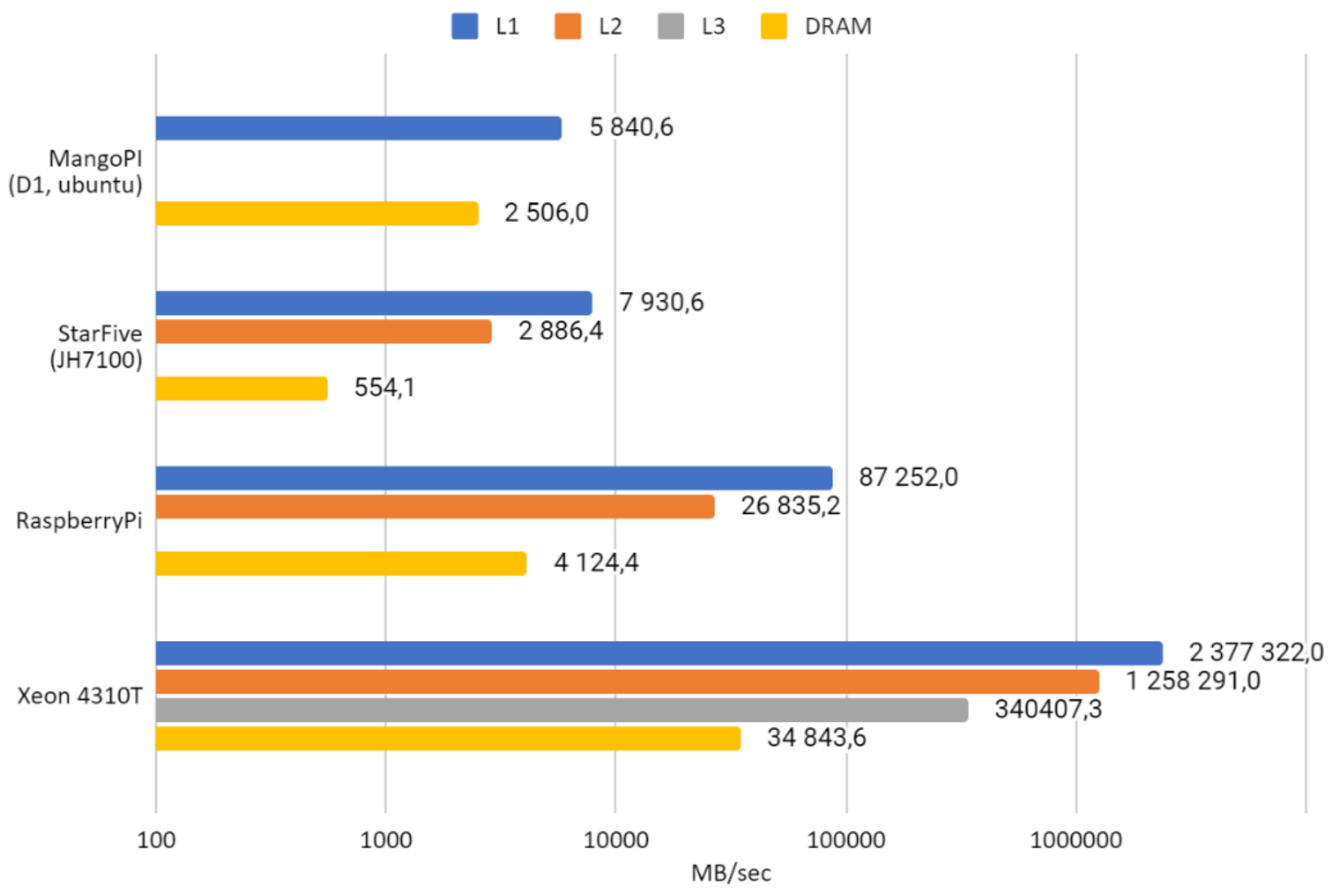

**Fig. 1.** The results of the STREAM benchmark

We select the sizes of the arrays in such a way that they are not forced out of the memory of the considered level and could not be cached efficiently in faster memory. All levels of memory that are available on each specific device are considered. We run a multi-threaded (for a shared memory) or sequential (for an individual resource, for example, an L1 cache) version of STREAM. In the sequential experiments the results are multiplied by the number of cores. Overall, we use the maximum value that is achieved during sufficiently large number of repetitions of the experiment.

The results of the obtained throughputs are presented in Fig.1. It turned out that the RISC-V devices have a number of drawbacks. We found that there is only L1 cache with a rather low bandwidth on the Mango Pi board with the Allwinner D1 processor compared to other devices. In the case of the StarFive VisionFive board on JH7100, we observe the low bandwidth of DRAM memory which corresponds to the reduced memory channel in the device. Overall, we can say that the memory subsystem of the considered RISC-V devices is significantly behind its analogue on ARM and, as expected, is even more inferior to the x86 Xeon CPU.

### 4.2 In-place Dense Matrix Transposition Algorithm

**Algorithm.** The in-place dense matrix transposition algorithm is one of the key algorithms in linear algebra [24]. It is used both as a standalone procedure and as part of other linear algebra algorithms. In this section, we consider a sequence of optimizations to the matrix transposition algorithm that incrementally improve performance from the most basic (naïve) implementation to the efficient high-performance version.

**Naïve Implementation (“Naïve”).** The following implementation (Listing 1) is a code that a programmer often develop without thinking about performance optimization. Of course, such an implementation cannot be expected to perform efficiently, but it is inefficient for all the devices. All are on equal terms.

```
Transpose_baseline (double * mat, int size)
   for (i=0; i < size; i++)
          for (j=i+1; j < size; j++)
                 mat[i][j] = mat[j][i]
```

**Listing 1.** Pseudocode of the naïve implementation

**Parallelization (“Parallel”).** Since most modern hardware is multi-core, using multithreading is an important way to reduce the computation time. For this algorithm, the parallel implementation almost does not differ from the reference one. Since the iterations of the outer loop are independent of each other, the algorithm can be easily parallelized using the OpenMP technology. Note that OpenMP is supported by all compilers on 4 considered devices. However, the Allwinner D1 (Mango Pi) device is single-core, so it makes no sense to use parallelism there, and other optimizations for this device are performed in sequential code.

**Better Data Reuse: Cache Blocking (“Blocking”).** The next optimization is to avoid unnecessary data loads and better reuse of data already loaded into caches. This can be achieved by block traversal of a matrix, which is typical for many matrix algorithms. Listing 2 shows the pseudocode without unnecessary implementation details.

```
Transpose_block (double * mat, int size)
  parallel_for (i_blk=0; i_blk<size; i_blk+=blk_size)
    for (j_blk=i_blk; j_blk<size; j_blk+=blk_size)
      for (i=i_blk; i<i_blk+blk_size; i++)
        for (j=j_blk +1; j<j_blk+blk_size; j++)
          mat[i][j] = mat[j][i]
```

**Listing 2.** Pseudocode of the block algorithm implementation

**Improved Memory Access (“Manual_blocking”).** The next optimization continues and enhances the ideas of the previous one. Its main task is to provide, if it is possible, sequential access to RAM. To do this, blocks are loaded into the cache manually, after which they are transposed and data is exchanged with other blocks. The pseudocode is shown in Listing 3.

```
Transpose_improvedMemAccess (double * mat, int size)
  parallel_for(i_blk=0; i_blk<size; i_blk+=blk_ size)
```

```
      double cache_blk[blk_size*blk_size]
      for (j_blk=i_blk; j_blk<size; j_blk+=blk_size)
        load_block_to_cache (i_blk, j_blk)
        transpose_block_in_cache()
        swap_block (j_blk, i_blk)
        transpose_block_in_cache()
        store_block (i_blk, j_blk)
```

**Listing 3.** Pseudocode of the improved block algorithm implementation

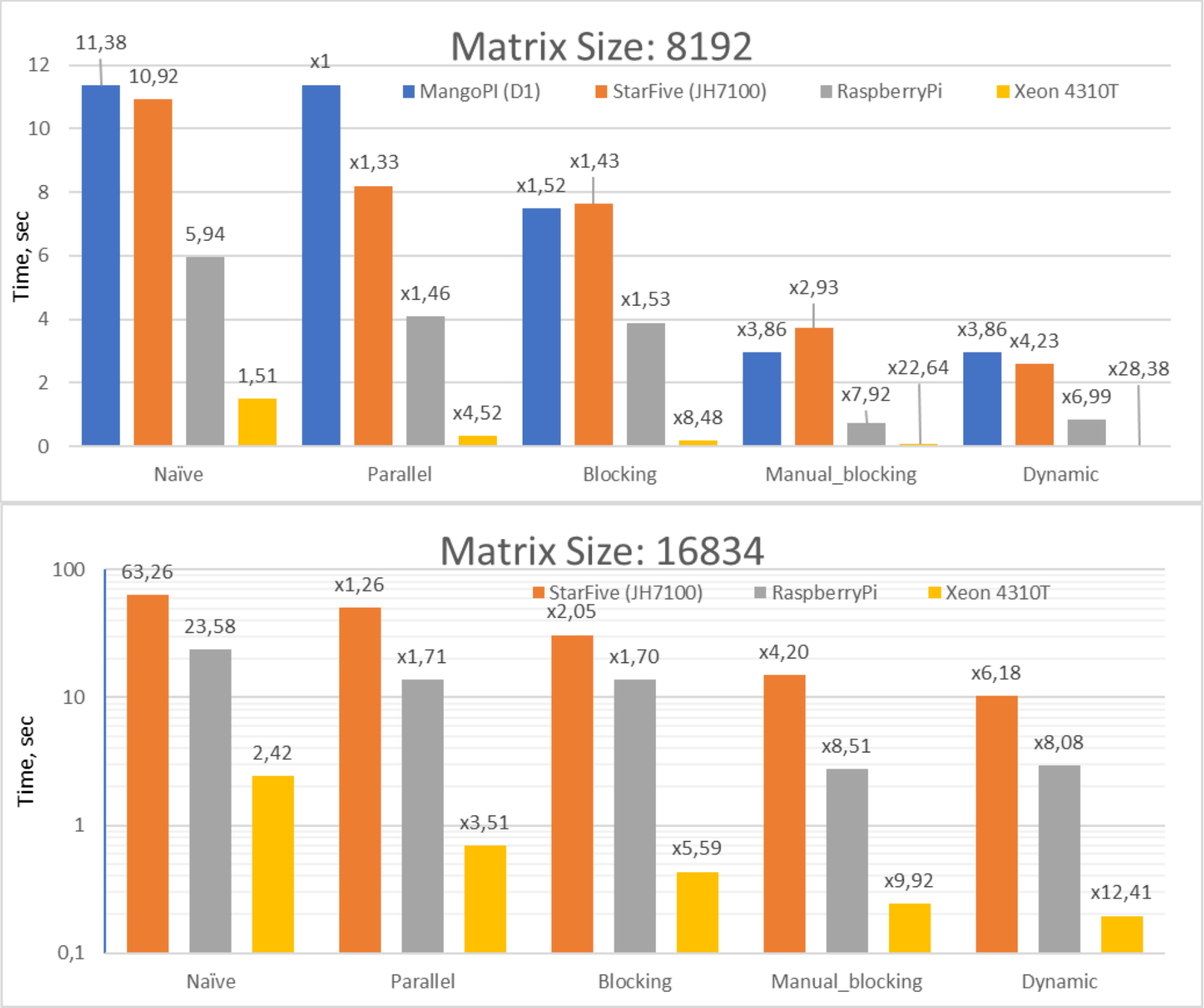

**Fig. 2.** Computation time of five implementations of the matrix transposition algorithm on four computing devices (Intel Xeon server, Raspberry Pi and two RISC-V boards). The labels above the bars of the diagram show the computation time of the naïve version of the algorithm given in seconds, as well as the speedup of the optimized implementations relative to the naïve ones on the corresponding devices. The top panel contains the results for a 8192 x 8192 matrix, the bottom one for a 16384 x 16384 matrix. The bottom panel does not contain results for Mango Pi because the matrix does not fit in memory of this device.

**Dynamic Scheduling ("Dynamic").** The final version of the code differs from the previous one in the dynamic scheduling of the parallel loop. It makes it possible to eliminate the imbalance in the computational load that occurs in traversing the rows of an upper triangular matrix, which obviously have different lengths.

**Performance Results and Discussion.** Fig. 2 shows the computation time of the presented algorithms on four devices. In accordance with the previously introduced metrics, it shows the computation time of the naïve version of the algorithm on different devices, as well as the acceleration of optimized implementations relative to the naive version for each of the platforms. The lack of acceleration of parallel implementations (Parallel and Dynamic versions) on Mango Pi is due to the single-core CPU.

We found that optimizations that were developed for the x86 architecture perform well also on RISC-V devices. Despite significant architectural differences between the devices, the memory subsystems are organized with the similar principles, so optimizations have made it possible to better utilize memory resources of RISC-V, ARM and x86 CPUs. Note that the presented optimizations show a good acceleration, especially considering that this algorithm does not use vector instructions, which in many cases can speed up calculations and make working with memory more efficient.

Given the substantially larger computing capabilities of Intel Xeon, we compare the overall computation time on RISC-V and Raspberry Pi CPUs. Note that despite the very large advantage of the latter in memory bandwidth at the STREAM benchmark over both RISC-V devices, the gap in computation time between RISC-V and ARM is much smaller. Moreover, with an increase of the matrix size to 16384, the difference in speedup compared to the naïve version on ARM and RISC-V CPUs decreases. It confirms better utilization of available resources of RISC-V CPU.

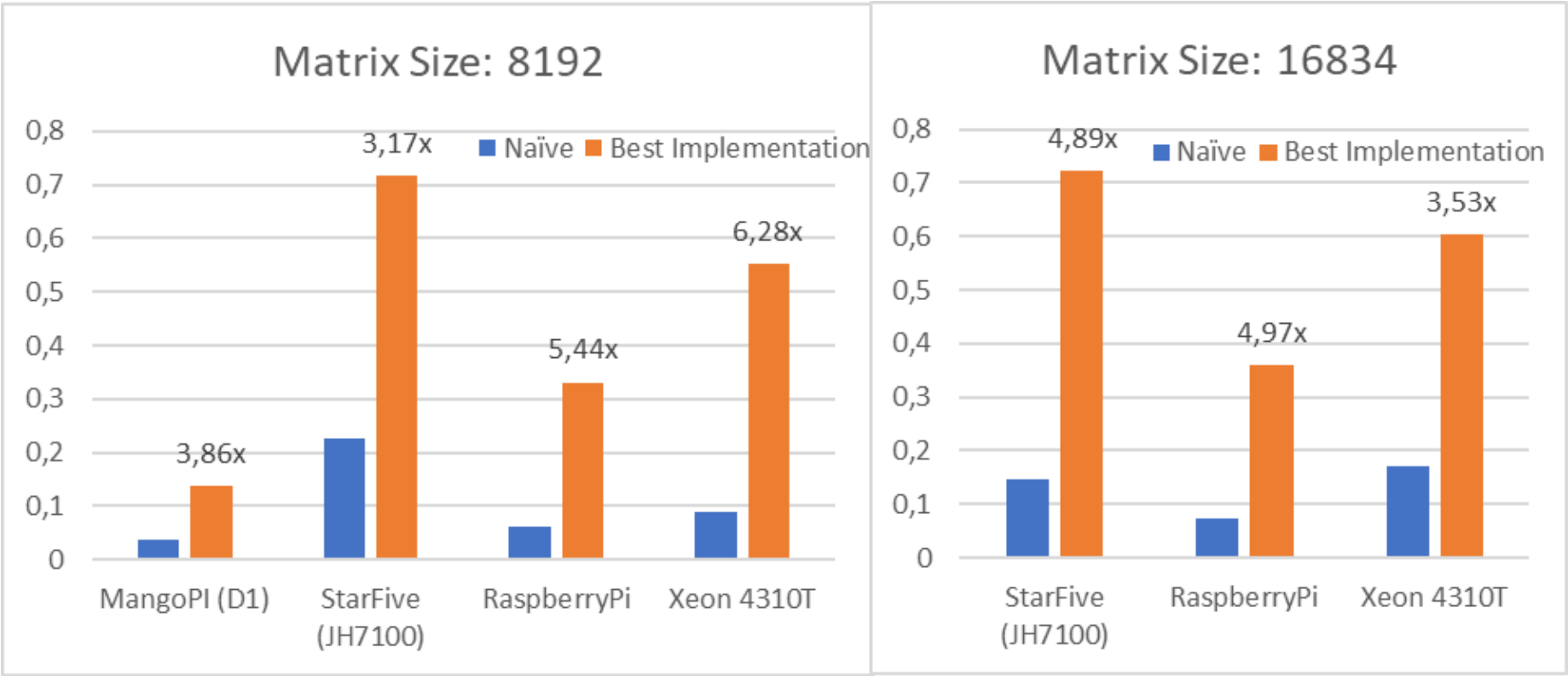

**Fig. 3.** Effectiveness of the relative memory bandwidth utilization for four devices. The metric is calculated for the naïve implementation and the best optimized implementation for each of the devices. The matrix size is 8192 x 8192 (left panel) and 16384 x 16384 (right panel). The right panel does not show results for the Mango Pi because the matrix does not fit in memory.

Comparing the results of two RISC-V boards with each other, we noticed that despite the good memory bandwidth on the device with the Allwinner D1 (Mango Pi) processor compared to the second one on the JH7100, their computation time is almost identical. To find out the reason for this phenomenon and analyze the performance in terms of one of the previously announced metrics, we calculated the relative memory bandwidth utilization (Fig. 3). This metric shows how efficiently the reuse of data loaded from memory is implemented, and how significantly the computation time depends on the properties of the memory subsystem. The optimal value of this metric, equal to one, is not achievable in many cases, but closeness to one indicates efficient memory utilization. All devices show almost the same increase in this relative indicator for sufficiently large matrices. In the case of Raspberry Pi, it seems unusual that memory utilization is at such a low level. Probably, this is due to the lack of ARM-specific optimizations, but in this case the devices are on equal terms, because we run C codes without architecture-specific optimizations only. The metrics show that StarFive (JH7100) performed well in terms of memory bandwidth utilization. This is primarily due to the low memory bandwidth on StarFive, however, this board has two memory channels for two cores. In the case of Mango Pi (D1), it can be seen that there is a low memory utilization both in the naïve implementation and in the most optimized one. Note that this device has only one level of cache with only modest improvements compared to DRAM, which affects the performance.

Summing up the results of optimizations of the in-place dense matrix transposition algorithm on RISC-V devices, we note that despite the expectedly large difference in the computation time, the available RISC-V CPUs make it possible to achieve a high degree of utilization of resources using commonly applied optimization techniques.

### 4.3 Gaussian Blur Algorithm

**Algorithm.** In this section we consider an image filtering with a Gaussian Blur algorithm as a benchmark. The problem is formulated as follows. Let there be an image (tensor) containing one or three channels at the input. Each image pixel contains one or three intensity values, respectively, each in the range from 0 to 255, or from 0 to 1, if normalization is performed. The problem of filtering involves passing through the image from left to right and from top to bottom, applying the Gaussian filter kernel to the pixels, and calculating a discrete convolution. The output is an image that has the same spatial dimensions as the input one and contains updated intensity values.

We chose the filtering task as a benchmark for the following reasons. Firstly, it is necessary in many computer vision (CV) algorithms for the preliminary preparation of input data. Secondly, there are efficient implementations of the Gaussian filter for different computing architectures, in particular, in OpenCV. Therefore, there are optimized implementations to compare performance. The third and most important reason is that discrete convolution is a basic operation of convolutional neural networks, which are commonly used in CV applications. The performance of convolutions significantly affects the time of a direct pass through the neural network, which is critical in the implementing deep neural network models in real applications. The efficiency of the implementation of this operation on the target hardware highly influences the overall computation time when solving CV problems using convolutional networks. Therefore, the filtering task is the first step towards deep neural networks inference optimization on RISC-V architectures. Then we consider several implementations of the algorithm that consistently improve the efficiency of utilization of computing resources, and, as before, analyze the achieved performance results.

**Naïve Implementation ("Naïve").** As a basic implementation, we use an algorithm in which the Gaussian filter kernel is used to sequentially calculate the intensities of each pixel of the resulting image row by row (Listing 4).

```
1 :  cntChannel = 3
2 :  middle = sizeFilter / 2
3 :  for (i = 0; i < h - sizeFilter; i++)
4 :    for (j = 0; j < w - sizeFilter; j++)
5 :      for (c = 0; c < cntChannel; c++)
6 :        sum = 0.f
7 :        for (i_f = 0; i_f < sizeFilter; i_f++)
8 :          for (j_f = 0; j_f < sizeFilter; j_f++)
9 :            pos_i = (i + i_f) * (w * cntChannel)
10:            pos_j = (j + j_f) * (cntChannel) + c
11:            sum += srcData[pos_i + pos_j] *
                 filter[i_f * sizeFilter + j_f]
12:        i_d = i + middle; j_d = j + middle
13:        distData[(i_d*w+j_d)*cntChannel + c] = sum
```

**Listing 4.** Pseudocode of the naïve implementation of the Gaussian Blur algorithm

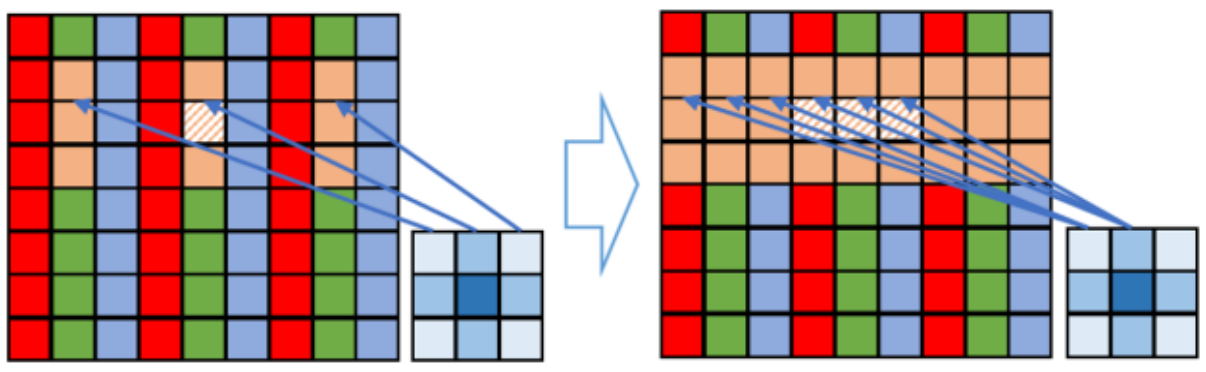

**Fig. 4.** Gaussian Blur filter optimization for color images: unit-stride memory access. Left panel: naïve implementation. Right panel: improved implementation.

**Unit-stride Access ("Unit-stride").** Note that if a color image is used, then memory access is not unit-stride (Fig. 4., left panel). As a first modification, we change the order of the loops so that the loop through the image channels (line 5) is inside the filter kernel application loop (line 8). As a result, memory access is unit-stride (Fig. 4., right panel), which is much better in terms of memory usage.

**One-dimensional Kernels ("1D_kernels").** For further optimization, we rearrange the computations [25] based on the following representation of the Gaussian filter:

$$G(x,y) = \frac{1}{2\pi\sigma^2} e^{-\frac{x^2+y^2}{2\sigma^2}} = \left(\frac{1}{\sqrt{2\pi}\sigma} e^{-\frac{x^2}{2\sigma^2}}\right)\left(\frac{1}{\sqrt{2\pi}\sigma} e^{-\frac{y^2}{2\sigma^2}}\right) \quad (1)$$

Now we can successively apply two one-dimensional Gaussian filter kernels instead of using a two-dimensional kernel (see Fig. 5).

The use of one-dimensional filters reduces the computational complexity of the entire algorithm. When using two-dimensional filters, the complexity can be estimated as $O(W \cdot H \cdot C \cdot F^2)$, where W and H are the width and height of the image, respec-tively, C is the number of channels, and F is the size of the filter kernel. By using two one-dimensional filters, the complexity can be reduced to $O(W \cdot H \cdot C \cdot F)$.

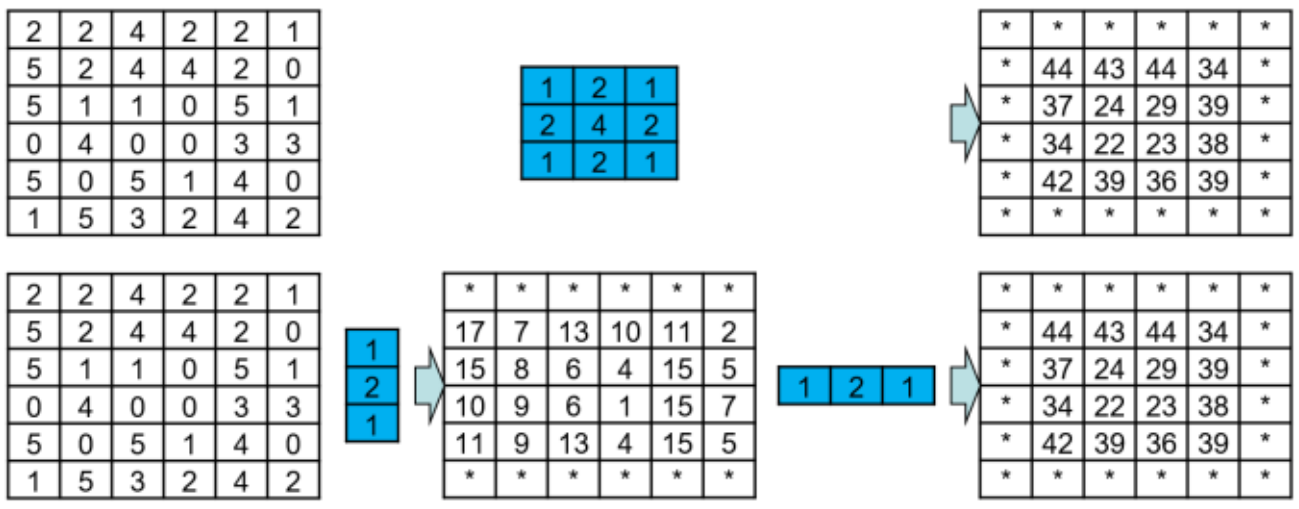

**Fig. 5.** Applying of the two-dimensional kernel (top row) and two one-dimensional kernels (bottom row) of the Gaussian blur filter

**Improving Memory Access ("Memory").** In the previous implementation, the performance of applying the horizontal kernel of the Gaussian filter is low due to an inefficient memory access pattern. Therefore, we use the order of loops, in which each element of the kernel interacts with the entire row from the original image matrix (Listing 5).

```
for (i = 0; i < h - sizeFilter; i++)
  for (i_f = 0; i_f < sizeFilter; i_f++)
    pos_i = (i + i_f) * (w * cntChannel)
    for (j = 0; j < w * cntChannel; j++)
      tmpData[(i + middle) * w * cntChannel + j] +=
         srcData[pos_i + j] * filter1D[i_f]
```

**Listing 5.** Pseudocode of the improved implementation of the Gaussian Blur algorithm

**Parallel Implementation ("Parellel").** The computations are independent and well-balanced, therefore we parallelize the algorithm trivially by using `#pragma parallel for` from OpenMP.

**Performance Results and Discussion.** We used a color image of size 2544 x 2027 for the experiments. To apply the filter, the intensities of each pixel were converted to the float type. The size of the kernel of the Gaussian filter $F = 19$. The computation time and the speedups achieved are shown in Fig. 6. As before, the sizes of the bars correspond to the computation time of a particular algorithm on the corresponding device, while the captions above the bars show the computation time for the naïve version and speedup of other versions relative to the naïve implementation. The experimental results allow us to draw the following conclusions. Firstly, we found that the baseline implementation lags behind OpenCV[1] by several orders of magnitude, regardless of the device architecture. Then, the computation time of the first modification of the algorithm ("Unit-stride") is obviously better because of sequential memory

[1] In the case of using processors with RISC-V architecture, the OpenCV computation time was measured on a Linux image that supports vector instructions

access which us much faster due to an efficient data prefetch. Apparently, this is not the case for the StarFive device, where data prefetch does not speed up calculations because low memory bandwidth does not allow data to be prepared on time and only leads to additional overhead.

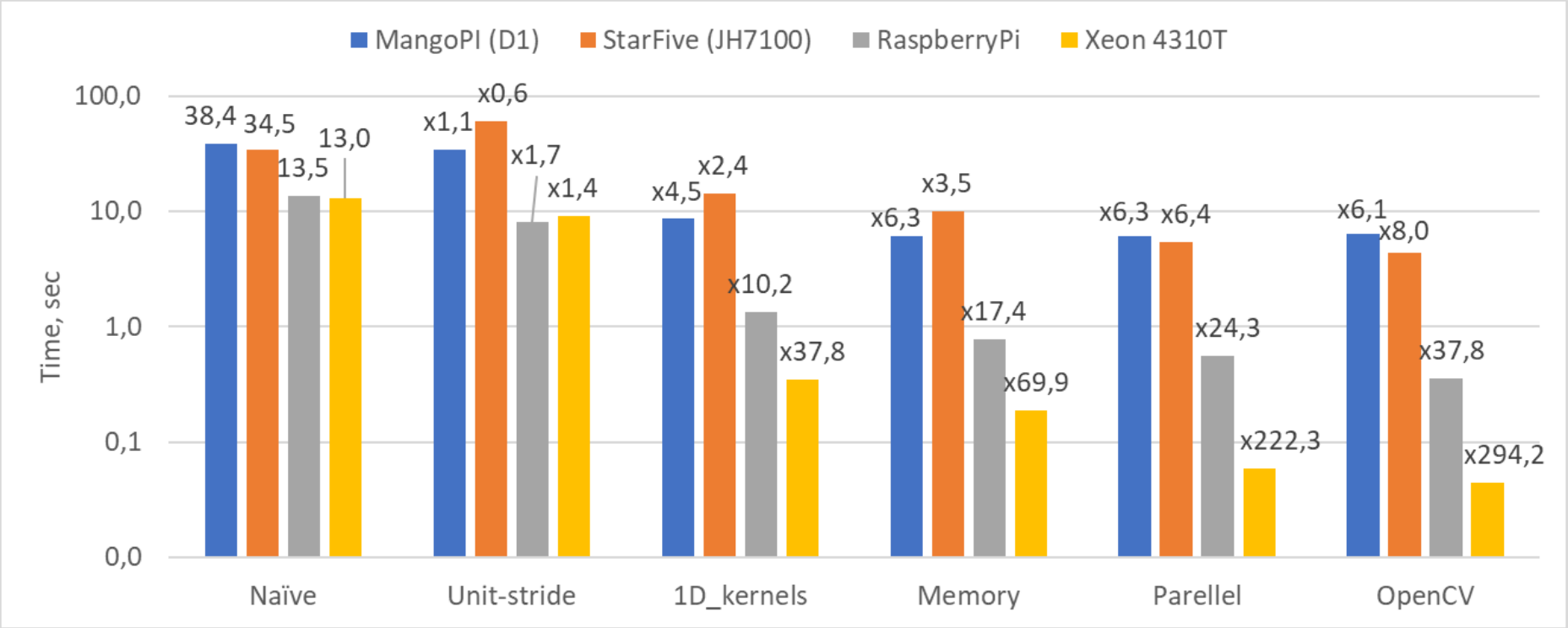

**Fig. 6.** Computation time of five implementations of the Gaussian Blur algorithm on four computing devices (Intel Xeon server, Raspberry Pi and two RISC-V boards). The labels above the bars of the diagram show the computation time of the naïve version of the algorithm given in seconds, as well as the speedup of the optimized implementations relative to the naïve ones on the corresponding devices.

Next, we paid attention to the computation time and speedup of the next modification of the algorithm ("1D_kernels"). As expected, the calculations are faster. It is worth noting in particular that with a filter size of $F = 19$, one would expect a substantial speedup. Apparently, it did not happen due to an excessive amount of memory access.This assumption is confirmed by the results of the following modification of the algorithm ("Memory"). Due to much more efficient memory access, the speedup compared to the naïve implementation becomes much larger. The acceleration by more than 19 times on the server with Intel Xeon 4310T processors is justified by the fact that the compiler has been able to vectorize the code with the loop order, used in the "Memory" implementation. The computation time and speedup of parallel implementations are shown in Fig. 6 ("Parallel" bar). Since the problem is memory bound, speedup is limited by the number of available memory channels.

Like for the matrix transposition benchmark, we analyze the effectiveness of memory bandwidth usage (see Fig. 7). When calculating this relative metric (see section 3.3), we used a "1D_kernels" implementation as a baseline. We conclude that the memory subsystem of Mango Pi does not allow for high performance of the image filtering algorithm due to the lack of L2 cache and slow L1 cache. StarFive lags behind RaspberryPi in memory access performance, but overall, the results are comparable. In case of Intel Xeon 4310T, the parallel algorithm provided an increase in the memory bandwidth usage metric due to the presence of a larger number of memory channels, which are not available in other devices under consideration.

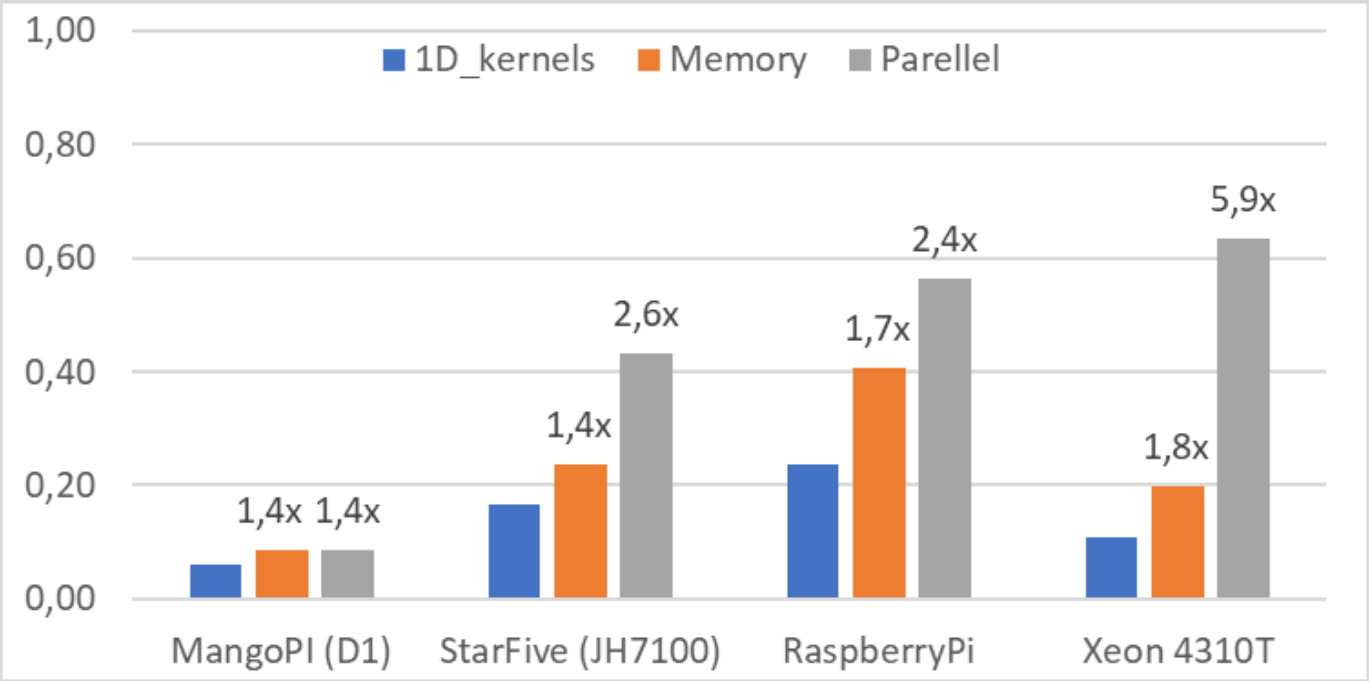

**Fig. 7.** Effectiveness of the relative memory bandwidth utilization for four devices. The metric is calculated for the three optimized implementations of the Gaussian Blur algorithm. Labels show the improvement compared to the "1D_kernels" implementation.

## 5 Conclusion

In this paper, we explored new opportunities and perspectives of the RISC-V computing architecture. Despite the many papers studying of architectural ideas and their possible implementations, testing existing RISC-V devices and studying their performance issues are of great interest. In this regard, the paper presents an analysis of the performance of two RISC-V devices on three memory-bound benchmarks in comparison with well-studied ARM and x86 CPUs. First, we measured the memory bandwidth using the commonly used STREAM benchmark. The results showed that the existing RISC-V prototypes are still significantly behind ARM and x86 devices. Therefore, when considering two benchmarks from linear algebra and image processing, we proposed to take into account not only the computation time, but also the efficiency of using the memory subsystem. It turned out that in the matrix transposition algorithm, RISC-V devices demonstrate excellent utilization of available resources, while when filtering images, memory is used less efficiently, which is caused by a small number of memory channels. We especially want to note that the typical memory optimization techniques worked out over the past decades on ARM and x86 behave as expected on RISC-V, allowing to significantly speed up calculations. As a result, we can conclude that although the available RISC-V devices are not yet suitable for HPC, but nevertheless they show a high potential for further development.

**Acknowledgements.** The study is supported by the Lobachevsky University ("Priority 2030" program). Experiments were performed on the Lobachevsky supercomputer.